\documentstyle[sprocl]{article}
\begin{document}

\title{Probing the Fifth State of Matter with Relativistic \\
Heavy Ions:  A Theoretical Overview}

\author{Berndt M\"uller}

\address{Department of Physics, Duke University, Durham, NC 27708-0305}

\maketitle\abstract{
I review the current status of lattice Monte-Carlo computations of the 
equation of state of QCD, our current understanding of the thermalization 
process at collider energies, and two new theoretical developments, 
one pertaining to the ab-initio calculation of the low-$x$ gluon structure 
of heavy nuclei, the other to the chaotic behavior of QCD.  At the end 
I give a brief overview of the status of various quark-gluon plasma 
signatures.}

\section{Introduction}

In the preceding talk, Johanna Stachel \cite{JS96} surveyed the status 
of relativistic heavy ion experiments with fixed targets at the
Brookhaven AGS and the CERN-SPS.  Whereas the observed hadron yields
and spectra provide strong evidence that the final state---at the
moment of break-up---is approximately thermal, we have little
information about when the thermalization is achieved during the
course of the nuclear reaction.  Since the energy deposition processes
at AGS and SPS energies are dominated by soft strong interactions, we
currently also lack the theoretical tools to make reliable
calculations.  Hence we do not know yet, and may never know, whether a
thermalized quark-gluon plasma is formed in these reactions.

The situation appears to be much more favorable at the higher energies
where the future heavy ion colliders, RHIC and LHC, will operate.  Two
developments over the past few years have raised the hope that
accurate calculations of the initial phase of nuclear reactions at
such energies will eventually be possible.  Firstly, numerical
simulations of lattice QCD are rapidly approaching the stage where
quantitatively reliable predictions of the equations of state of
strongly interacting matter and its dynamical properties will be
feasible.  Secondly, the inclusion of color screening into parton
cascades and nuclear structure functions opens up the prospect of a
seamless description of the reaction dynamics within the framework of
perturbative QCD from the initial nuclear structure up to a thermalized 
quark-gluon plasma.  If this expectation comes to fruition, it will be 
possible to test quantitative theoretical predictions at RHIC and at the LHC.

In this talk I will review the current status of lattice Monte-Carlo
computations of the equation of state of QCD, our current
understanding of the thermalization process at collider energies, and
two recent new theoretical developments, one pertaining to the
ab-initio calculation of the low-$x$ gluon structure of heavy nuclei,
the other to the chaotic behavior of QCD.  At the end I will give a
brief overview of the status of various quark-gluon plasma signatures.

\section{The QCD Phase Diagram}

Rigorous results about the phase diagram of QCD are based on
Monte-Carlo simulations of lattice-QCD at finite temperature.  For the
pure SU(3)-gauge theory, quantitatively reliable results including an
extrapolation to the continuum limit have recently become available.
Calculations by the Bielefeld group \cite{Boy96} on lattices up to 
$32^3\times 8$ yield a critical temperature (in units of the string 
tension $\sigma$) of
\begin{equation}
T_c/\sqrt{\sigma} = 0.629 \pm 0.003,
\end{equation}
or $T_c=260$ MeV if the value of $\sigma$ derived from the Regge slope
is used.  The phase transition is of first order with a latent heat of
slightly less than $2T_c^4$.  The simulations provide clear evidence
of significant interactions extending far above $T_c$.  Fitting to
predictions of thermal perturbation theory one finds, e.g.,
\begin{equation}
g^2 (5T_c) \approx 1.5.
\end{equation}
The simulations also provide information about quantities of dynamical
relevance, such as the speed of sound, which is found rising from
$c_s^2 <0.1$ near $T_c$ to $c_s^2 \approx 1/3$ at $T=5T_c$.

Simulations of full lattice-QCD with light dynamical quarks have not
yet reached this advanced stage, but have also made significant progress.  
State-of-the-art calculations \cite{Ber96} show a steep jump in the
quantity $\epsilon/T^4$ at $T_c\approx 150-200$ MeV, indicating a
rapid unthawing of new degrees of freedom.  The uncertainty in the
value of $T_c$ is mainly caused by discrepancies between different
ways of setting the scale (string tension or hadron masses).
Although it remains unclear whether thermodynamic quantities will show
a singularity at $T_c$ in the continuum limit and what order phase
transition it may be, the simulations clearly demonstrate that the
quark condensate $\langle \bar\psi\psi\rangle$ drops steeply at
exactly the same temperature:  color deconfinement and chiral symmetry
restoration occur at the same temperature.  The results are expected
to reach a similar quality as those presently existing for the pure
gauge theory when the next generation of parallel computers with close
to teraflops performance will become available.

Perturbative approaches, valid far above $T_c$, have also made
significant progress.  The equation of state is now known \cite{AZ95,ZK95}
up to order $g^5$, and a scheme yielding a complete perturbative result 
with only minimal lattice input has been worked out \cite{BN95}.  The 
``bad news'' is that it is now understood that the convergence of the 
perturbative series (even in the sense of an asymptotic series)
requires $g\le 1$ or $\alpha_s(T) \le 0.1$, which only holds for
temperatures above the electroweak unification scale \cite{BN96}.

Nevertheless, thermal perturbation theory provides important insight
into the dynamics of the high-temperature phase of QCD:
\begin{enumerate}

\item quarks and gluons develop dynamical masses of order $gT$.

\item Long-range color forces (except static color magnetism) are
dynamically screened on distances of order $(gT)^{-1}$, providing an
effective infrared cut-off for many transport coefficients.

\item Colored particles propagating through the plasma constantly
exchange their color with the medium, rendering the quark-gluon plasma
a poor color conductor.  As will be discussed below, this ``color
chaos'' facilitates the rapid equilibration of degrees of freedom in
the plasma.
\end{enumerate}

\section{Initial conditions at RHIC and LHC}

Most recent theoretical predictions for the initial conditions at
which a thermalized quark-gluon plasma will be produced at heavy ion
colliders are based on the concept of perturbative partonic cascades.
The parton cascade model \cite{GM92} starts from a relativistic transport
equation of the form
\begin{equation}
p^{\mu}{\partial\over\partial x^{\mu}}F_i(x,p) = C_i(x,p\vert F_k)
\end{equation}
where $F_i(x,p)$ denote the phase space distributions of QCD quanta.
The collision terms $C_i$ are obtained in the framework of
perturbative QCD from elementary $2 \to 2$ scattering amplitudes
allowing for additional initial- and final state radiation due to
scale evolution of the perturbative quanta.  To regulate infrared
divergences, the parton cascade model requires a cut-off for the $2 \to
2$ scattering (usually $p_T^{\rm min}=1.5 - 2$ GeV/$c$) and a cut-off
for time-like branchings $(\mu_0^2 = 0.5 - 1$ GeV$^2/c^2)$.

Numerical simulations of such cascades for heavy nuclei provide a
scenario where a dense plasma of (predominantly) gluons develops in
the central rapidity region between the two colliding nuclei shortly
after the impact.  Detailed studies \cite{EW94} show that the momentum 
spectrum of partons becomes isotropic and exponential, i.e. practically 
thermal, at a time $\tau\approx 0.7 \Delta z$ in the rest frame of a 
slab of width $\Delta z$ at central rapidity.  To allow for a hydrodynamic
description, the width of the slab should exceed the mean free path of
a parton.  Including color screening effects, one finds that the mean
free path of a gluon in a thermalized plasma is $\lambda_f\approx
(3\alpha_s T)^{-1}$ where $T$ is the thermal slope of the parton
spectrum.  For the very high initial values $(T\ge 0.7$ GeV) one
concludes that a thermal hydrodynamic picture makes sense after
$\tau_i\approx 0.3$ fm/$c$.

The high density of scattered partons in $A + A$ collisions makes it
possible to replace the arbitrary infrared cut-off parameters
$p_T^{\rm min}$ and $\mu_0^2$ by dynamically calculated medium-induced
cut-offs \cite{Bir93}.  The dynamical density-dependent screening of 
color forces eliminates the need for $p_T^{\rm min}$, and the suppression 
of radiative processes provided by the Landau-Pomeranchuk-Migdal effect
makes the virtuality cut-off $\mu_0^2$ superfluous.  Note that the
viability of this concept crucially depends on the high parton
density:  the dynamical cut-off parameters must lie in the range of
applicability of perturbative QCD.  Since the density of initially
scattered partons grows as $(A_1A_2)^{1/3}(\ln s)^2$, this condition
requires both large nuclei and high collision energy.  The
calculations indicate that this criterion will be met at RHIC and LHC
but not at the presently accessible energies of the SPS and AGS.
The framework is also not applicable to $pp$ or $p\bar p$ collisions
at current energies because the parton density remains too low.

The dynamic screening of parton cascades can be implemented as
follows \cite{EMW95}.  According to the uncertainty principle, a 
parton-parton collision can be considered as complete after a 
``formation time'' $\tau_f\approx \hbar/p_T$ in the c.m. frame, where 
$p_T$ is the momentum transfer.  Accordingly, harder collisions are 
completed first.  One can then consider the hard collisions as 
effectively screening the softer ones.  The screening mass applicable 
to parton collisions of scale $p_T$ is then determined as
\begin{equation}
\mu^2(p_T) = -{3\over\pi^2} \alpha_s(p_T^2) \int_{p_T}^{\infty} d^3k
\vert\nabla_k \; f(k) \vert
\end{equation}
where $f(k)$ is the momentum density of more violently scattered
partons.  Feeding $\mu^2(p_T)$ back into the differential cross
sections determining the number of scatterings one obtains a
differential equation for $f(k)$ or, equivalently, $\mu^2(p_T)$.  For
Au + Au collisions at RHIC energy, the screening mass saturates at
slightly below 1 GeV at small $p_T$, and at 1.5 GeV for Pb + Pb
collisions at the LHC.  Both these values are comfortably within the
range of applicability of perturbative QCD, demonstrating that there
may be no need for an artificial infrared cut-off.  The screening of
parton scattering by already scattered partons is analogous to the
interaction among ladders in the traditional picture of soft hadronic
interactions \cite{Mat76}.  It would be interesting to rederive 
these results from this alternative point of view.

The self-consistent parton cascade makes parameter-free (though \break
somewhat model dependent) predictions about the initial conditions
achieved at RHIC and LHC.  For the heaviest nuclei one expects
thermalization to occur at $T=730$ MeV (RHIC) or $T_c$ = 1150 MeV
(LHC) and initial energy densities of about 60 GeV/fm$^3$ (RHIC) or
430 GeV/fm$^3$ (LHC).  At the initial moment $(\tau\approx 0.25$
fm/$c$) the parton plasma is not chemically equilibrated.  It takes
another several fm/$c$ to achieve chemical equilibrium. Following the
evolution with the hydrodynamical model including self-consistent
screening, the quark-gluon plasma is expected to last for about 5
fm/$c$ until the temperature falls to $T_c$ at RHIC and somewhat
longer at the LHC.  Final multiplicities are predicted to be
$dN/dy\approx 1700$ (RHIC) and $\approx 8000$ (LHC).

\section{New Theoretical Approaches}

\subsection{Semiclassical parton structure}

While representing a major advance in the application of perturbative
QCD to relativisitic nuclear collisions, the self-screened parton
cascade still relies on the input of experimentally measured parton
structure functions of the colliding nuclei.  A new approach, due to
the Minneapolis group \cite{MV94} promises to make these structure 
functions themselves calculable within perturbative QCD.  The basic idea
underlying this approach is that, as seen by partons with $x \le
10^{-2}$, the valence quarks in a heavy nucleus constitute a very
dense, sheet-like random color source.  The large area density
$\rho\approx 3A/\pi R^2$ defines a large scale parameter $\mu^2=\rho$
at which the QCD coupling $\alpha_s(\mu^2)$ is weak.  Introducing
light-cone coordinates it is then possible to formulate a systematic 
program for calculating gluon and quark (sea) structure functions at 
small $x$ as generated by classical random color fields and their quantum
fluctuations.  At the classical level, corresponding to the
Weizs\"acker-Williams approximation, the gluon density for not too
small $p_T$ is given by
\begin{equation}{1\over \pi R^2}\; {dN\over dxdp_T^2} = {\alpha_s\mu^2
(N_c^2-1) \over \pi^2 x p_T^2}.
\end{equation}
Loop corrections will introduce $(\ln x)/x$ contributions, possibly
leading to a power-like behavior at small $x$ upon resummation.

The approach can be extended to the collision between two nuclei,
viewed as the interaction among two counter-propagating sheets of
valence quarks \cite{KMW95}.  Since screening effects are already 
included in this approach, it may provide another description of the 
initial state produced in a nuclear collision at RHIC or LHC.
\vfill
\eject

\section{Chaotic Glue and Thermalization}

A completely different approach to the thermalization problem in
non-Abelian gauge theories emerges from solutions of the clasical
Yang-Mills equations.  Numerical calculations have shown that these
equations form a strongly chaotic, infinite-dimensional dynamical
system \cite{MT92,Go94}.  The evidence is derived from the study of the 
(properly defined) distance between two initially neighboring field
configurations as a function of time.  Neighboring configurations
diverge at an exponential rate:
\begin{equation}
\Vert A_1^{\mu}(t) - A_2^{\mu}(t)\Vert \sim e^{\lambda t}.
\end{equation}
The complete spectrum of Lyapunov exponents $\lambda_i$ can be
obtained from numerical solutions of the Yang-Mills equations on a
lattice in Minkowski space.  There is evidence that the lattice fields
thermalize (in the microcanonical sense) and the Lyapunov exponents
scale as $(g^2T)$.

A fundamental result of nonlinear dynamics is that the sum of all
positive Lyapunov exponents (the so-called Kolmogorov-Sinai entropy)
measures the rate of entropy growth after coarse graining.  The
Lyapunov exponents, therefore, provide a measure of the thermal
equilibration time.  Using the largest Lyapunov exponent for the SU(3)
gauge theory \cite{Go93}, one finds a characteristic time of order 
0.2 - 0.3 fm/$c$ in the range of temperatures (300-700 MeV) relevant for 
RHIC.  This agrees nicely with the results from parton cascade simulations.

\section{Quark-Gluon Plasma Signatures} 

A wide variety of signatures for the formation of a quark-gluon plasma 
in heavy-ion collisions has been proposed.  Considerable progress has 
been made in recent years in understanding the background to many of 
these signals \cite{Mu95,HM96}.  For example, it will be very difficult, 
if not impossible, to identify direct lepton pairs emitted by the plasma, 
because of a formidable background of decay leptons from D-mesons 
\cite{Gav96}.  On the positive side, the suppression of charmonium states 
observed in $p+A$ collisions is becoming understood as due to the 
absorption of the color octet component in nuclei \cite{KS96}.  This 
provides a well-defined benchmark against which additional suppression 
effects can be identified.  Indeed, the very high initial temperature now 
predicted for RHIC and the LHC will allow even charmed quarks to get into 
chemical equilibrium before hadronization.  There is also a chance that 
even $\Upsilon$ mesons will be significantly suppressed at the LHC.
The predicted production of multistrange baryons above thermal yields 
also remains a promising signature.

It has recently been understood that the energy loss of hard partons
in a quark-gluon plasma may be much higher than expected \cite{Bai95}.  
It may even grow with the length of the traversed medium.  Accordingly, 
the quenching of jet production, or of leading high-$p_T$ hadrons, has
attracted increased interest as a probe of the early stage of the
evolution of the plasma.

The formation of chirally disorientated domains of the quark
condensate in the vacuum is a new, intriguing signature of the chiral
phase transition \cite{RW93}.  Such domains correspond to coherent 
excitations of the pion field; they would decay by producing large 
fluctuations away from 1/3 in the $\pi^0/\pi$ ratio.  Detailed numerical 
studies \cite{AHW95} of the linear sigma model have shown how disorientated 
domains can grow if the chiral transition occurs rapidly, producing a 
temporarily unstable state of the chiral order parameter 
$\langle\bar\psi\psi\rangle$.  Sophisticated techniques for the 
identification of domain structures, such as wavelet analysis \cite{Hua96}, 
may help observing these domains if they are produced experimentally.

\section{Summary}

As RHIC proceeds toward completion, theoretical advances during the
last few years are giving us a much clearer view of the physics to be
expected in heavy ion collisions at that energy.  Lattice-QCD
simulations have unambiguously established the rapid ``unthawing''
of color and the restoration of chiral symmetry at a temperature below
200 MeV.  Calculations of parton transport processes based on
perturbative QCD predict very high initial temperatures, above 500 MeV
at RHIC and 1 GeV at the LHC.  New approaches to low-$x$ parton
structure and the thermalization problem hold the promise of a model
and parameter-independent description of the formation and evolution
of a quark-gluon plasma at RHIC energy and beyond.  Finally, the
various plasma signatures are becoming much better understood,
providing valuable guidance for the experimental program at RHIC which
will start in 1999.

\section*{Acknowledgments}

This work was supported in part by a grant from the
U.S. Department of Energy, Office of Energy Research (DE-FG02-96ER40945).
The author also thanks the organizers of the PANIC-96 conference for
support.

\end{document}